# SASE FEL Storage Ring

G. Dattoli, E. Di Palma, A. Petralia and J. V. Rau

*Abstract*—We explore the possibility of operating a SASE FEL with a Storage Ring. We use a semi-analytical model to obtain the evolution inside the undulator by taking into account the interplay on the laser dynamics due to the induced energy spread and to the radiation damping. We obtain the Renieri's limit for the stationary output power and discuss the possibility of including in our model the effect of the beam instabilities.

## I. INTRODUCTION

FREE Electron Lasers (FEL) operating with Storage Rings (SR) have attracted much interest since the first year of FEL Physics. They have been the topic of intensive studies both theoretical and experimental. The Renieri's limit [1], later confirmed by the experiment in ACO [2], has been the most notable result of the earliest theoretical investigations of SR-FEL.

According to Renieri, the average power attainable with a SR-FEL oscillator is proportional to the power lost via synchrotron radiation in a machine turn.

The ACO experiment [2] displayed other important features, associated with the laser dynamics, in particular, the interplay between FEL induced energy spread and damping, leading to the equilibrium configuration, prompted Ellaume [3] to provide a very transparent model combining the FEL and SR dynamics, modelled through a set of rate equations, yielding the turn by turn evolution of the laser intensity and of the induced energy spread. Subsequent investigations [4] clarified the role of the pulse propagation in SR-FELs, providing a general picture of the spatial and spectral properties of the laser radiation. The SUPER-ACO (SA) FEL experiment has been the source of other important observation, regarding the interplay of the laser dynamics with e-beam instabilities, like head tail and saw tooth [5]. An extension of the Elleaume model, put forward in ref. [6], allowed the understanding of the conditions permitting the suppression of the laser instabilities at the onset of the FEL operation [7]-[8]. The further analysis of the SA-FEL experiment revealed that the laser operation allowed an increase of the Touschek life-time [5]. In ref. [9], the use of the model described in refs. [3]-[6], implemented with the inclusion of the increase of the beam transverse dimensions induced by the FEL interaction, provided the theoretical understanding of these effects.

More recent results obtained at ELETTRA [10] and UVSOR-II [11] have further proved the flexibility of the SR-FEL device to study the nonlinear harmonic generation mechanisms, which confirmed the power scaling relations predicted in ref. [12].

The set of experiments performed in [10]-[11] should be considered as "single pass" experiment. The harmonic generation is, indeed, induced by an external seed and the repetition rate of the seed is such that the damping restores the natural spread of the e-beam, which can be considered "fresh" after each interaction.

In this paper, we will consider the case of a SAFER (acronym of SASE-FEL-SR), namely a FEL operating in the mirror-less configuration, using a storage ring as beam source, and we will follow the turn by turn evolution by combining the FEL high gain equations and the SR energy spread rate equations.

It should be mentioned that this type of device has been originally discussed in ref. [13]. Here we will treat the problem from a different point of view and show that the presence of a FEL SASE section on a storage ring, if not exploited to produce radiation, may be an efficient tool to cure the machine instabilities.

## II. THE MODEL

In this section, we will introduce the model equations and the relevant numerical handling. We will use an Elleaume type modelling of the SR-FEL rate equations by following the lines developed in refs. [3]-[6] whose main tools have been:

A reliable parameterization of the FEL oscillator equation, with the inclusion of the gain dependence on bunch length and energy spread.

Coupling of the FEL oscillator rate equation to the turn by turn SR beam dynamics, accounting for the FEL induced energy spread and consequent damping due to synchrotron radiation emission in the ring bending magnets.

There is a substantive difference between SR-SASE and SR-Oscillators FELs. In the latter, the radiation is not stored in an optical cavity, but the power growth occurs in one pass of the beam through the undulator. The feedback between FEL radiation and the ring is provided by the FEL induced energy spread and damping, which only partially restores the initial beam quality, when the electrons enter the undulator for the successive iteration.

The strategy we will follow is that of using the SASE FEL high gain equations reported in refs.[3],[6] with the SR-damping rate equation for the SR. We remind that the power

G. Dattoli is with ENEA C.R. Frascati, Frascati (RM), Italy (corresponding author to provide phone: +39-06-94005421; e-mail: Giuseppe.Dattoli@enea.it).
E. Di Palma, A. Petralia are with ENEA C.R. Frascati, Frascati (RM), Italy
J. V. Rau is with Istituto di Struttura della Materia, Consiglio Nazionale delle Ricerche, Via del Fosso del Cavaliere, 100 - 00133 Rome, Italy.

growth along the undulator and the induced energy spread are provided by the following relations [12]:

$$P(z) = \frac{P_0}{9} \frac{A(z)}{1 + \frac{P_0}{9P_F}[A(z)-1]}$$

$$A(z) = 3 + 2\cosh\left(\frac{z}{L_g}\right) + 4\cos\left(\frac{\sqrt{3}}{2}\frac{z}{L_g}\right)\cosh\left(\frac{z}{2L_g}\right)$$

$$L_g = \frac{\lambda_u}{4\pi\sqrt{3}\rho}, P_F = \sqrt{2}\rho P_E$$

$$\sigma_i^2(z) \cong \frac{9}{4}\frac{\rho}{P_E}\frac{P_F - P_0}{P_F - P(z) + 1.24[P(z)-P_0]}P(z)$$

(1)

where $P_E$ is the seed power and $P_F$ the saturated power, in a single passage.

The turn by turn damping for the energy spread can be written as:

$$\sigma_{n+1}^2 = (1 - 2\frac{T}{\tau_s})[\sigma_n^2 + \sigma_{i,n}^2]$$ (2)

where $T = L_R/c$, $L_R$ is the ring circumference, $\tau_s$ the damping time and $\sigma_{i,n}^2$ is the energy spread induced by the FEL interaction at the n-th iteration.

The power growth will be affected by the induced energy spread because:

i) the gain length increases with energy spread increase, according to:

$$L_g = L_g^{(0)}\left(1 + 0.185\sqrt{\frac{3}{2}}\tilde{\mu}^2\right) \qquad \tilde{\mu} = \frac{2\sigma_T}{\rho}$$

$$\sigma_T = \sigma_{\varepsilon,0}\left(1 + \left(\frac{\sigma}{\sigma_{\varepsilon,0}}\right)^2\right)^{\frac{1}{2}}$$

(3)

where $L_g^{(0)}$ and $\sigma_{\varepsilon,0}$ are respectively the gain length in the homogeneous regime and the beam energy spread at the first passage ($n=0$ and $\sigma_0=0$);

ii) the $\rho$ parameter decreases, since the increase of the energy spread determines a corresponding increase of the bunch length, thus providing a reduction of the peak current, according to the following identity:

$$\hat{I} = \frac{\hat{I}^{(0)}}{\left(1 + \left(\frac{\sigma}{\sigma_{\varepsilon,0}}\right)^2\right)^{\frac{1}{2}}}$$ (4)

Therefore, since $\rho \propto \hat{I}^{1/3}$, we obtain

$$\rho = \rho^{(0)}\left(1 + \left(\frac{\sigma}{\sigma_{\varepsilon,0}}\right)^2\right)^{-\frac{1}{6}}$$ (5)

Putting all these things together, we obtain the gain length at the n-th pass as:

$$L_g^{(n)} = L_g^{(0)}\left(1 + \left(\frac{\sigma_n}{\sigma_{\varepsilon,0}}\right)^2\right)^{\frac{1}{6}}\left[1 + 0.185 \cdot \frac{\sqrt{3}}{2}\tilde{\mu}_0^2\left(1 + \left(\frac{\sigma_n}{\sigma_{\varepsilon,0}}\right)^2\right)^{\frac{4}{3}}\right]$$ (6)

$$\Pi(z,\sigma_n^2) = \frac{P_0}{9}\frac{A(z,\sigma_n^2)}{1 + \frac{P_0}{9P_F}[A(z,\sigma_n^2)-1]}$$

$$\frac{d}{dt}\sigma^2 = -\frac{2}{\tau_s}(\sigma^2 - \sigma_i^2)$$

$$\frac{d}{dt}P = -\frac{1}{\tau_s}P + \Pi(Z_s,\sigma^2)$$

(7)

$$Z_s = 1.066 L_g^{(0)} \ln\left(\frac{9P_F}{P_0}\right)$$

where we have transformed the difference equation into a differential equation according to the prescription:

$$A_{n+1} - A_n \cong T\frac{d}{dt}A$$ (8)

The undulator length $Z_s$ corresponds to that allowing full saturation for the fresh bunch operation. The results of a preliminary simulation are shown in Fig. 1, where we have reported:
1) the turn by turn induced energy spread;
2) the turn by turn output power;
3) the power growth inside the undulator at different turn.

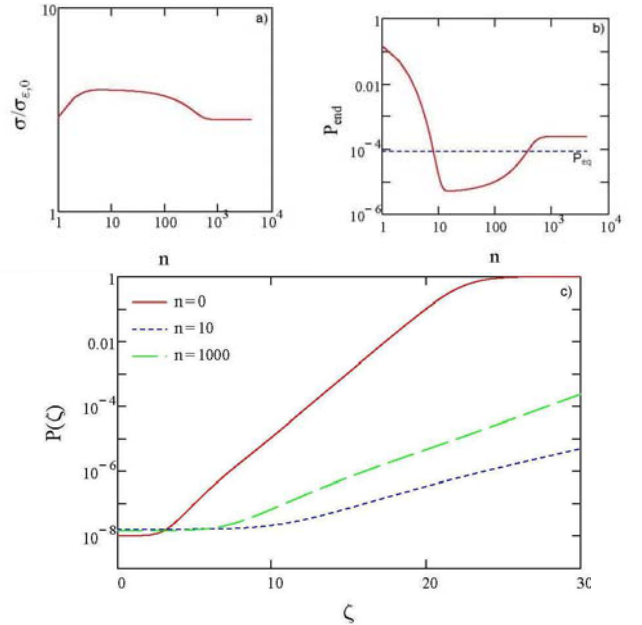

Fig. 1. SAFER dynamics, with an undulator corresponding to 30 gain length: (a) Induced energy spread $(\sigma/\sigma_{\varepsilon,0})^2$; (b) SASE power normalized to $P_F$, the dash line represents the equilibrium power also normalized to $P_F$, $P_{eq} = P_e/P_F$; (c) Power growth Vs $\zeta = Z/L_g^{(0)}$, at different iterations: Continuous line n=1, Dot line n=10, Dash line n=1000. The total number of iterations corresponds to 3 damping times.

The previous results have been obtained by assuming

$$\frac{\tau_s}{T} \cong 2.5 \cdot 10^3$$

compatible with a Storage Ring having a damping time of 10 ms and a Ring circumference of 0.75 Km. The Physics underlying the process described in Figs.1 is fairly transparent: at the first iteration the system reaches the full saturation, with a consequent significant increase of the energy spread. At the successive steps the output power decreases because of the largest undulator length necessary to reach the full saturation. The output power is switched off, the induced spread is damped, and the laser process can start again and eventually reaches a kind of equilibrium, which albeit far from saturation, is significantly above the spontaneous emission level (of a factor $10^3$-$10^4$).

Such a behaviour is not dissimilar from what has been discussed for the SR-Oscillator FEL dynamics.

The equilibrium power can be estimated from the equation:

$$P_e \cong \frac{2}{9}\frac{T}{\tau_s}\tilde{\mu}_0^2 \rho_0 P_E \qquad (9)$$

which can be easily derived from the first of eq. (7) by imposing $d\sigma^2/dt=0$ and by imposing the condition that, at the equilibrium, the power is much less than the saturated power[1].

Multiplying both sides by the electron bunch duration ($\sigma_E/c$) we get:

$$\overline{P}_e = \chi P_S, \chi = \frac{8}{9}\frac{\sigma_0^2}{\rho_0}$$
$$\overline{P}_e = \frac{\sigma_E}{cT}P_e \qquad (10)$$
$$P_S = \frac{NeE_E}{\tau_s}$$

which is just a restatement of the Renieri limit, with N the number of particles in the bunch and $E_E$ the electron beam energy. We have, indeed, obtained that also in a SAFER device the average laser power is proportional to the synchrotron radiation power emitted in one machine turn.

### III. SAFER DYNAMICS AND TOUSCHECK EFFECT

In the previous discussion, we have stressed that the saturation of a SAFER occurs through the induced energy spread, which determines:
1) a reduction of the $\rho$ parameter through the decrease of the peak current,
2) an increase of the gain length.

Fig. 1 is relevant to an operation, in which the undulators length is chosen to be long enough to guarantee the full saturation in the first passage, when the beam has not undergone any FEL induced diffusion. This is not mandatory, the undulator can be shorter than the length, corresponding to a fresh beam saturation. In Fig. 2, we report the same as in Fig. 1, but for a shorter undulator.

Although the equilibrium power is slightly lower than the full length undulator, this configuration provides the emission of a substantive amount of power during the transition to the equilibrium configuration. In the first of refs. [14] a behaviour of the type reported in Figs. (1,2) has been recognized as "partial lasing".

In the quoted papers [14]-[16] the Authors consider the potential of an ultimate storage ring for future light sources and consider different advanced facilities under commissioning, construction or study. They make explicit reference to PEP-X study ref. [16] which foresees a ring with a circumference of 2 km long and having a peak current of several hundred of Ampère. The length of the ring is such that long undulators can be accommodated on the machine circumference straight section. The simulations performed in [14] yields undulator length of the order of 100 m. Furthermore their partial lasing power growth agrees with our semi-analytical results.

The analysis we have developed so far is essentially based on a phenomenological model, which neglects some features and therefore our approximations deserve further comments.

In particular we did not consider the interplay of the FEL dynamics with the SR longitudinal dynamics. A correct description would require the use of the Fokker Planck equation to study the evolution of the longitudinal distribution including the different mechanisms associated with FEL diffusion, quantum diffusion, FEL induced variations of the partition numbers, violation of the Robinson theorem in the second of ref. [2] such a treatment has been accomplished for SR FEL oscillators. In refs.[17] a specific analysis was addressed to the effect of FEL (amplifier) dynamics on the electron beam longitudinal damping times. This study has put in evidence that the damping times of higher order longitudinal modes decrease with the FEL intensity, therefore ensuring a faster damping of these modes. Although we did not repeat the simulations for the case discussed in the present article we do not expect deviations from such a general trend.

A further point is the distortion of the longitudinal phase space induced by the FEL interaction, we have included this effect in the evaluation of the turn by turn induced energy spread and bunch lengthening. We have however implicitly assumed that the phase space distribution remains Gaussian albeit with a larger longitudinal phase space emittance. This can be not true and higher order modes, yielding a phase space distortion, may be induced; according to the previous statement we expect that they are efficiently suppressed by the larger damping times, characterizing the mode. It may be however argued that being the number of turns quite large the longitudinal phase space can be distorted by competition of the FEL and SR longitudinal dynamics. We can exclude significant effects, since, as shown in the second of refs. [17], for these values of the laser power intensity the longitudinal phase space dynamics is dominated by a diffusive behavior, which does not induce significant filamentation.

The interaction of the FEL with the ring environment does not occur through the induced energy spread and the consequent damping. A fairly accurate description of the new elements, affecting the FEL dynamics in the presence of a FEL type interaction, has been provided in refs. [7],[10], where the consequence of the beam induced heating and the relevant feedback on Saw-tooth type instabilities [7],[8], Touschek intra-beam scattering [9] and Touschek life time have been studied.

---

[1] The last of eqs. (1) yields for $P(z) \ll P_F$ the following approximation for the induced energy spread $\sigma_i^2(z) \cong \frac{9}{4}\frac{\rho}{P_E}P$

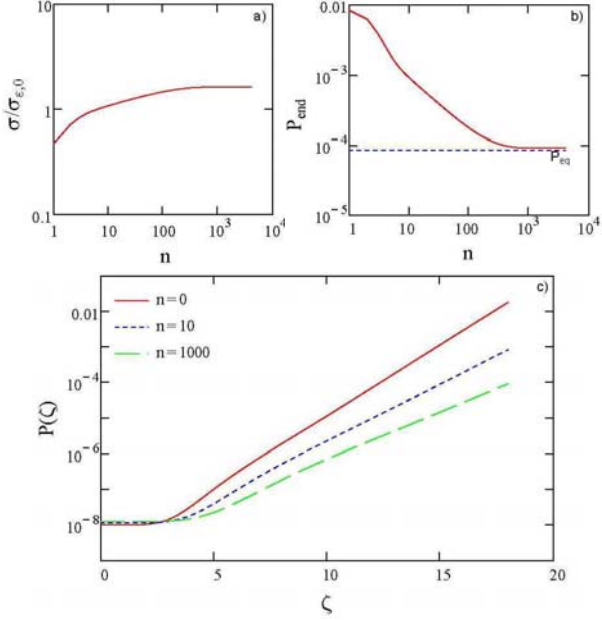

Fig. 2. Same as Fig. 1, but for a shorter undulator, corresponding to 18 gain length.

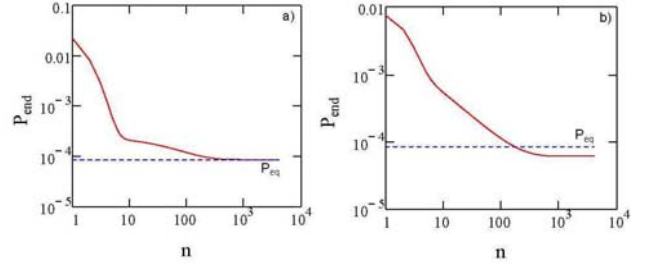

Fig. 3. Same as Fig. 1b), but for 25 gain length long undulator: (a) $C_x=2$, (b) $C_x=6$

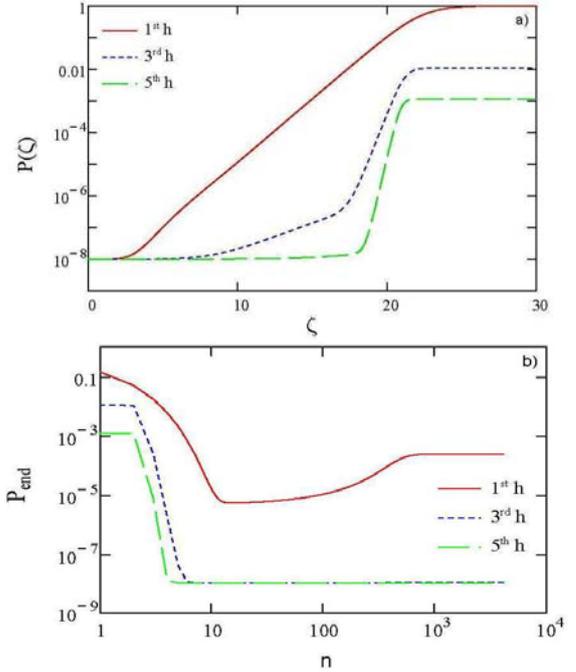

Fig. 4. SAFER operation at higher harmonics: (a) fundamental and higher harmonics; (b) maximum power vs. the number of turns for the fundamental and the higher harmonics.

The undulator section and the rest of the ring are not separated environments and further element causing a reduction of the $\rho$ parameter is the presence of a non-zero dispersion function ($\eta_{x,y}$), which determines a transverse-longitudinal coupling, causing a dependence of the e-beam transverse section on the energy spread, according to:

$$\sigma_{x,y} = \sqrt{\beta_{x,y}\varepsilon_{x,y} + (\eta_{x,y}\sigma_\varepsilon)^2} \qquad (11)$$

where $\beta_{x,y}$ and $\epsilon_{x,y}$ are the Twiss parameters (radial and vertical) and emittances, respectively. As a consequence, the induced energy spread is responsible for an increase of the beam dimensions, with a consequent further increase of the gain length due to a reduction of the $\rho$ parameter, namely:

$$\rho \propto J^{\frac{1}{3}} \qquad J = \frac{\hat{I}_p}{2\pi\sigma_x\sigma_y} \qquad (12)$$

where we have defined the current density $J$.

Assuming that the radial part only of the off axis function is non-zero, we write

$$\rho = \rho^{(0)}\left[\left(1+\left(\frac{\sigma}{\sigma_0}\right)^2\right)\left(1+C_x\left(1+\left(\frac{\sigma}{\sigma_0}\right)^2\right)\right)\right]^{-\frac{1}{6}} \qquad (13)$$

$$C_x = \left(\frac{\eta_x\sigma_0}{\sqrt{\beta_x\varepsilon_x}}\right)^2$$

In Fig. 3, we have reported the turn by turn evolution of a SAFER with different values of the horizontal off energy function.

It is evident that a more appropriate analysis requires a specific choice of a machine lattice, including an analysis of magnetization errors, and an appropriate maximization of the dynamic aperture, for efficient injection and maximization of the beam life-time. This goes beyond the goal of the paper and will not be discussed.

## IV. CONCLUDING COMMENTS

The analysis of the dynamics should be completed with the inclusion of the interplay of SASE FEL with the various instabilities, affecting a ring, and what are the conditions allowing the SAFER operation in their presence, or even if, as stressed in previous papers [7]-[9], the onset of the laser operation may mitigate their effect on the stability of the machine. This is an extremely interesting problem, which may involve a new role of SASE FEL on storage rings as an optional insertion device, for the control of the instabilities which may occur in Storage Rings with extremely bright electron beams as those considered in refs. [15],[16]. In this case the mechanism inhibiting the instability growth is in between the effect induced by a storage ring FEL discussed in ref. [8] and that of a damping wiggler [18]. The equilibrium power necessary to switch off the instability can be calculated by requiring that the induced FEL energy spread be

comparable with that due to the instability itself, which is given by [8]

$$\sigma^* \cong \sqrt{\delta^{3/2} - 1}\, \sigma_0$$
$$\delta = \frac{I}{I_{th}} \quad (14)$$

With $I$ being the bunch current and $I_{th}$ the threshold current, above which the instability manifests itself [19]. From eq. (10) we find that the FEL average power sufficient to damp the instability is

$$P_e \cong K \frac{T}{\tau_s} \tilde{\mu}_0^2 P_F \quad \text{with} \quad K = \frac{\sqrt{2}}{9} \frac{1}{\rho} \left( \delta^{\frac{3}{2}} - 1 \right) \frac{T}{\tau_s} \tilde{\mu}_0^2 \quad (15)$$

Since $K$ may be of the order of $10^{-3}$, for properly chosen parameters, the "partial lasing" underlying the process, we have described can be, in principle, exploited to damp the instability.

Different solutions can however be adopted including a FEL oscillator or an amplifier using an external laser. The oscillator option is rather difficult to be implemented because we are considering high energy accelerators and problems arise with the optics due either to the absence of appropriate mirrors for the radiation confinement and to their degradation due to the intense x-ray flux they had to sustain.

The heater solution with an external laser could be an interesting tool, which might reduce the length of the undulator section, the only drawback being the use of an external element to the device.

A more appropriate comparison between the various solutions will be discussed elsewhere.

Before closing the paper, we want to analyze the possibility that the system is capable of providing coherent power due to nonlinear harmonic generation mechanism, which should occur at least during the first passages. In Fig. 4, we report the growth of the higher order harmonics. As expected, during the first few turns the induced energy spread has not reached sufficiently large values to inhibit the growth of the higher harmonics.

REFERENCES

[1] A. Renieri, "Storage ring operation of the free-electron laser: The amplifier", *Nuovo Cimento B*, vol. 53, pp. 160-178, 1979.
 G. Dattoli and A. Renieri, "The free-electron laser single-particle multimode classical theory ", *Nuovo Cimento B*, vol. 61, pp. 153-180, 1981.
[2] M. Billardon, P. Elleaume, J.M. Ortega, C. Bazin, M.Bergher, M. Velghe, D.A.G. Deacon, Y. Petroff, K.E. Robinson, and J.M.J. Madey, "First operation of a storage-ring free-electron laser", *Phys. Rev. Lett.*, vol. 51, pp. 1652-1655, 1983.
 see also G.N. Kulipanov, I.V. Pinaev, V.M. Popik, A.N.Skrinsky, A.S. Sokolov and N.A. Vinokurov, "The VEPP-3 storage-ring optical klystron: Lasing in the visible and ultraviolet regions", *Nucl. Instrum. Meth.*, vol. A296, pp. 1-3, 1990.
[3] P. Elleaume, "Macrotemporal structure of storage ring free electron laser", *J. Physique,* vol. 45, pp. 997-1001, 1984.
[4] P. Elleaume, "Microtemporal and spectral structure of storage ring free-electron lasers", IEEE J. Quantum Electron., vol. QE-21, pp. 1012-1022, 1985.
[5] J. J. Le Duff, "Current and Current Density Limitations in Existing Electron Storage Rings", *Nucl. Instrum. Meth.,* vol. A239, pp. 83-101, 1985.
[6] G. Dattoli, L. Giannessi, P. L. Ottaviani, A. Renieri, "A model for the saturation of a storage ring free electron laser", *Nucl. Instrum. Meth.,* vol. A365, pp. 559-563, 1995.
[7] G. Dattoli, L. Mezi, A. Renieri, M. Migliorati, M. E. Couprie, R. Roux, D. Nutarelli, and M. Billardon, "Storage-ring free-electron-laser dynamics and head-tail instability", *Phys. Rev. E.,* vol. 58, pp. 6570-6574, 1998.
[8] R. Bartolini, G. Dattoli, L. Mezi, A. Renieri, M. Migliorati, M. E. Couprie, G. De Ninno, R. Roux, "Suppression of the Sawtooth Instability in a Storage Ring by Free-Electron Laser: An Example of Nonlinear Stabilization by Noise", *Phys. Rev. Lett.*, vol. 87, pp. 134801-4, 2001.
[9] R. Bartolini, G. Dattoli, L. Giannessi, L. Mezi, A. Renieri, M. Migliorati, C. Bruni, M. E. Couprie, D. Garzella, G. Orlandi, "Saturation and electron-beam lifetime in a storage ring free-electron laser", *Phys. Rev. E,* vol. 69, pp. 036501(1-5), 2004.
[10] G. De Ninno, M. Trovò, M. Danailov, M. Marsi, E. Karantzoulis, B. Diviacco, R.P. Walker, R. Bartolini, G. Dattoli, L. Giannessi, L. Mezi, M.E. Couprie, A. Gatto, N. Kaiser, S. Günster, D. Ristau, "The UV European FEL at ELETTRA: towards compatibility of storage ring operation for FEL and synchrotron radiation", *Nucl. Instrum. Meth.,* vol. A 507, pp. 274-280, 2003.
[11] T. Tanikawa, G. Lambert, T. Hara, M. Labat, Y. Tanaka, M. Yabashi, O. Chubar, M. E. Couprie, "Nonlinear harmonic generation in a free-electron laser seeded with high harmonic radiation", *EPL*, vol. 94, pp. 34001, 2011.
[12] Dattoli G., Ottaviani P. L. and Pagnutti S., "Nonlinear harmonic generation in high-gain free-electron lasers", *J. Appl. Phys.,* vol. 97, pp. 113102(1-7), 2005.
[13] Z. Huang, K. Bane, Y. Cai, A. Chao, R. Hettel, C. Pellegrini, "Steady-state analysis of short-wavelenght, high-gain FELs in a large storage ring", *Nucl. Instr. Meth.,* vol. A 593, pp. 120-124, 2008.
[14] Y. Jing, Design of an ultimate Storage Ring, Thesis Indiana University (August 2011).
 P. Elleaume, A. Ropert, *Nucl. Instr. Meth.,*"The ultimate hard X-ray storage-ring-basedlight source", vol. A 500, pp- 18-24, 2003.
 M. Borland, "A super-bright storage ring alternative to an energy recovery linac", *Nucl. Instr. Meth.,* vol. A 557, pp. 230-235, 2006.
 K. Tsumaki, N. Kumagai, "Very low emittance light source storage ring", *Nucl. Instr. Meth.,* vol. A 565, pp. 394-405, 2006.
[15] M. Bei et al., "The Potential of an Ultimate Storage Ring for Future Light Sources", *Nucl. Instr. Meth.,* vol. A 622, pp. 518-535, 2010.
[16] R.Hettel et al., "Concepts for the PEP-X light source", in: Proceedings of the Particle Accelerator Conference, Vancouver, 2009.
[17] G. Dattoli, L. Mezi, P. L. Ottaviani, A. Renieri, and M. Vaccari," Storage-ring FEL amplifiers and electron beam longitudinal mode-damping times", Phys. Rev. E, vol. 57, pp. 7153-7161, 1998.
 G Dattoli,L Mezi," The storage-ring FOKKER–PLANCK equation: inclusion of the free electron laser amplifier contributions", Nucl. Instr. Meth., vol. A 436, pp. 415-429, 1999.
[18] M. Borland, L. Emery, "Direct space-charge calculation in ELEGANT and its application to the ILC damping ring", Proceedings of the PAC07 conference, 1124 (Albuquerque, New Mexico, 2007).
[19] J. L. Laclare, CERN Accelerator School, CERN 85-19, p. 377, 1985